      [1999/12/01 v1.4c Il Nuovo Cimento]
\documentclass{cimento}

\newcommand{\fermilat}{\emph{Fermi}-LAT}
\newcommand{\text}[1]{{\rm #1}}
\newcommand{\eqref}[1]{(\ref{#1})}

\newcommand{\valf}{v_\text{Alf}}

\newcommand{\be}{\begin{equation}}
\newcommand{\ee}{\end{equation}}

\newcommand{\galprop}{GALPROP}

\newcommand{\gray}{$\gamma$-ray}

\def\Dxx{D_{xx}}

\newcommand{\hi}{H~{\sc i}}

\usepackage{graphicx}  % got figures? uncomment this
\title{GALPROP: modeling cosmic ray propagation and associated\\ interstellar emissions}
\author{I.~V.~Moskalenko%\from{ins:x}%\ETC,
}
\instlist{
\inst{}%ins:x} 
Hansen Experimental Physics Laboratory and Kavli Institute for Astroparticle Physics and\\ Cosmology,
Stanford University\-Stanford, CA 94305, U.S.A
}
\PACSes{
\PACSit{95.35.+d}{Dark matter (stellar, interstellar, galactic, and cosmological)}
\PACSit{98.38.-j}{Interstellar medium (ISM) and nebulae in Milky Way}
\PACSit{98.70.Rz}{\gray{} sources; \gray{} bursts}
\PACSit{98.70.Sa}{Cosmic rays (including sources, origin, acceleration, and interactions)}
}
\begin{document}

\maketitle

\begin{abstract}
Research in many areas of modern physics and astrophysics such as, 
e.g., indirect searches for dark matter (DM), particle acceleration
in SNR shocks, and the spectrum and origin of extragalactic \gray{} background, rely heavily on studies of cosmic rays (CRs) and 
associated diffuse emissions.
New or improved instrumentation to explore these open issues is ready or under development. A fleet of
ground-based, balloon-borne, and spacecraft instruments measures many CR species, \gray{s}, 
radio, and synchrotron emission.
Exploiting the data collected by
the scientific missions to the fullest requires reliable and detailed calculations using a numerical model.
\galprop{} is the current state-of-the-art numerical CR propagation code
that has become a standard analysis tool in CR and diffuse \gray{} research.
It uses astrophysical information, nuclear and particle data as 
input to self-consistently predict CRs, \gray{s}, synchrotron emission and other observables.
This paper reviews recent \galprop{} developments and results.
%\texttt{cimento} class.
\end{abstract}

\section{The \galprop{} code}
%#########################################################################

The \galprop{} project~\cite{MS1998,SM1998} began in late 1996 and
has now 15 years of development behind it\footnote{http://sciencewatch.com/dr/erf/2009/09octerf/09octerfStronET/}. 
The code, originally written in fortran90, was made public in 1998. 
A version rewritten in C++ was produced in 2001, and the most recent public version 54 was recently released~\cite{GalpropWebrun}.
The code is available from the dedicated website\footnote{http://galprop.stanford.edu} where a facility for users to run the 
code via online forms in a web-browser is also 
provided. The key concept underlying the \galprop{} code is that various 
kinds of data, e.g., direct CR measurements including primary and secondary nuclei, 
electrons and positrons, \gray{s}, synchrotron radiation, and so forth, are all related to the 
same astrophysical components of the Galaxy and hence have to be modeled self-consistently~\cite{MSR1998}.
The goal is for GALPROP-based models to be as realistic as possible and to make use of available astronomical 
information, nuclear and particle data, with a minimum of simplifying assumptions.
A complete description of the rationale and motivation is given in the review~\cite{SMP2007}.
A very short summary of \galprop{} is provided below; for details the reader is referred to the relevant 
papers~\cite{MS1998,SM1998,GalpropWebrun,MS2000,Moskalenko2002,Moskalenko2003,SMR2000,Strong2004,Ptuskin2006}.

The \galprop{} code solves the CR transport equation with a given source 
distribution and boundary conditions for all CR species~\cite{SM1998}. 
This includes a galactic wind (convection), diffusive reacceleration in the
ISM, energy losses, nuclear fragmentation, 
radioactive decay, and production of secondary particles and isotopes.
The distribution of CR sources can be specified as required (\cite{Abdo2010cas} shows an example of the 
source distribution derived from the fit to the \fermilat{} data).
The numerical solution of the transport equation is
based on a Crank-Nicholson implicit second-order scheme~\cite{Press1992}. 
The spatial boundary conditions assume free particle escape.
For a given halo size the diffusion coefficient, as a function of
momentum and the reacceleration or convection parameters, is determined
from secondary/primary ratios. 
If reacceleration is included, the momentum-space diffusion
coefficient $D_{pp}$ is related to the spatial coefficient $\Dxx$ 
($= \beta D_0\rho^{\delta}$)~\cite{Seo1994},
where $\delta=1/3$ for a Kolmogorov spectrum
of interstellar turbulence or $\delta=1/2$ for a Kraichnan 
cascade (but can also be arbitrary), $\rho \equiv pc/Ze$ is the magnetic 
rigidity.
Non-linear wave damping~\cite{Ptuskin2006} can also be included if required.

Cross-sections are based on the extensive LANL database,
nuclear codes, and parameterizations~\cite{Mashnik2004}. 
The most important isotopic production cross-sections
are calculated using our fits to major production channels~\cite{Moskalenko2003,MM2003}.
Other cross-sections are computed using
phenomenological approximations~\cite{Webber2003} 
and/or~\cite{Silberberg1998} renormalized to
the data where they exist. 
The nuclear reaction network is built using the
Nuclear Data Sheets.
Production of neutral pions, secondary positrons and electrons is calculated 
using the formalism~\cite{Dermer1986a,Dermer1986b}
as described in~\cite{MS1998} 
with a correction from~\cite{Kelner2006} or 
using a parameterization given in~\cite{Kamae2005}.
Antiproton production uses formalism described in~\cite{Moskalenko2002}.

The \galprop{} code
computes a complete network of primary, secondary and tertiary CR
production starting from input source abundances.
Starting with the heaviest primary nucleus considered (e.g.\
$^{64}$Ni) the propagation solution is used to compute the source term
for its spallation products $A-1$, $A-2$ and so forth, which are 
then propagated in turn, and so
on down to protons, secondary electrons and positrons, and
antiprotons.  
To account for some special $\beta^-$-decay cases (e.g.,
$^{10}$Be$\to^{10}$B) the whole loop is repeated twice. 
The inelastically scattered protons and antiprotons are treated 
as separate components (secondary protons, tertiary antiprotons).
\galprop{} includes K-capture and electron stripping processes as 
well as knock-on electrons.

The \gray{s} are calculated using the propagated CR distributions, 
including a contribution from secondary
particles such as positrons and electrons from inelastic processes in the ISM 
that increases the \gray{} flux at MeV energies~\cite{Porter2008,SMR2004}.
The inverse Compton (IC) scattering is treated using the appropriate formalism for an
anisotropic radiation field~\cite{MS2000} 
%and employs
with
the full spatial and angular distribution of the interstellar radiation field (ISRF)~\cite{PS2005,Porter2008}. 
Electron bremsstrahlung cross section is calculated as described in~\cite{SMR2000}.
Gas-related \gray{} intensities ($\pi^0$-decay, bremsstrahlung) 
are computed from the emissivities as a
function of $(R,z,E_\gamma)$ using the column densities of \hi\ and
H$_2$ for Galactocentric annuli based on recent 21-cm and CO survey data with 
a more accurate assignment of the gas to the Galactocentric 
rings than earlier versions.
The synchrotron emission is computed using
a parameterization of the Galactic magnetic field.
The line-of-sight integration of the corresponding emissivities with the 
distributions of gas, ISRF,
and magnetic field yields \gray{} and synchrotron sky maps.
Spectra of all species on the chosen grid and the \gray{}
and synchrotron sky maps are output in standard astronomical 
formats for direct comparison
with data, e.g., FITS\footnote{http://fits.gsfc.nasa.gov/},
HEALPix\footnote{http://healpix.jpl.nasa.gov}~\cite{Gorski2005}, 
\emph{Fermi}-LAT MapCube format for use with LAT Science Tools 
software\footnote{http://fermi.gsfc.nasa.gov/ssc/data/analysis}, etc.

Also included in \galprop{} are specialized routines to calculate the propagation
of DM annihilation or decay products and associated diffuse \gray{} emission and synchrotron sky maps. 
The routines allow the DM profile, branching ratios, 
and particle spectra to be user-defined and calculate the source functions of the products of
DM annihilation and \gray{} emissivity. The particles are 
then propagated as separate species 
with the same propagation parameters as other CRs. 
The sky maps are calculated using the line-of-sight integration of the corresponding emissivities. 

Details of the
optimization of the code, linking to other codes 
(e.g., DarkSUSY~\cite{DarkSUSY2004,DarkSUSY2005}, SuperBayeS~\cite{Ruiz2006,Trotta2008a}) and so forth, 
can be found at the aforementioned website.

\section{Cosmic-ray propagation in the Galaxy}\label{CR_propagation}

Modeling CR propagation in the ISM includes the solution of the partial 
differential 
equation describing the transport 
with a given source distribution and boundary conditions for all CR species.
The diffusion-convection equation, sometimes incorporating diffusive 
reacceleration in the ISM, 
is used for the transport process and has proven to be remarkably successful 
despite its relative simplicity. 

Measurements of stable and radioactive secondary CR nuclei yield 
the basic information necessary to probe large-scale Galactic
properties, such as the diffusion coefficient and halo size, the
Alfv\'en  velocity and/or the convection velocity, as well as
the mechanisms and sites of CR acceleration.
Stable secondary
CR nuclei (e.g., $_{5}$B) can be used to determine ratio of halo size to the 
diffusion coefficient, while
the observed abundance of radioactive CR isotopes 
($^{10}_{4}$Be, $^{26}_{13}$Al, $^{36}_{17}$Cl, $^{54}_{25}$Mn) allows
the separate determination of halo size and diffusion coefficient~\cite{SM1998,Ptuskin1998,Webber1998,MMS2001}.
However, the interpretation of the sharp peaks observed in the 
secondary to primary CR
nuclei ratios 
(e.g., $_5$B/$_6$C, [$_{21}$Sc+$_{22}$Ti+$_{23}$V]/$_{26}$Fe) 
at relatively low energies, $\sim 1$-few GeV/nucleon, is model-dependent.

Closely connected with the CR propagation is the production of the
Galactic diffuse \gray{} emission, which is comprised of three
components: $\pi^0$-decay, bremsstrahlung, and IC.
Since the \gray{s} are undeflected by magnetic fields and 
absorption in the ISM is 
negligible~\cite{MPS2006}, they provide the information necessary 
to directly probe CR spectra and intensities in distant 
locations,  see \cite{MSR2004} for a review.
However, the interpretation of such observations is complicated 
since the observed \gray{} 
intensities are the line-of-sight integral of a sum of the three 
components of the diffuse
Galactic \gray{} emission, an isotropic 
component (often described as ``extragalactic''),
resolved and unresolved sources, together with instrumental background(s). 
Proper modeling of the diffuse \gray{} emission, including the 
disentanglement of the different components, requires well developed 
models for the ISRF and gas densities, together with the 
CR propagation~\cite{SMR2000,SMR2004}. 
Secondary CR particles and diffuse \gray{s} produced in conventional astrophysical 
processes constitute a background for potential exotic signals (e.g., from DM). 

For details of CR production and propagation the reader is referred to a recent review~\cite{SMP2007}.
A comprehensive summary of the indirect DM searches in CR and \gray{s} can be found
in~\cite{Porter2011}.

\section{Recent results} 
%#########################################################################

\subsection{Diffuse Galactic and isotropic \gray{} emission} 
%#########################################################################

The puzzling ``GeV excess" relative to the predictions of diffuse \gray{} emission models
based on locally measured CR spectra~\cite{SMR2000, Hunter1997} was an anomalous signal
observed in EGRET data above $\sim$1 GeV. 
It was proposed that the GeV excess results from annihilating
DM~\cite{deBoer2005}. This received much attention, but a number of
conventional explanations were also considered such as, e.g.,
variations in the CR spectra~\cite{MSR1998,SMR2004}. Paper \cite{Moskalenko2007} discusses the 
sources of systematic uncertainties in the EGRET calibration, data
handling, and in models of the diffuse emission. 

Testing the origin of the GeV excess was one of the early studies of the diffuse
\gray{} emission by the \fermilat{} team~\cite{Abdo2009diffuse111}. The data
at intermediate Galactic latitudes ($10^\circ< |b|<20^\circ$) were used in the study
because 
the diffuse \gray{} emission
over this region of the sky comes predominantly from relatively nearby CR nuclei
interactions with interstellar gas.
The \fermilat{} spectrum 
is well reproduced by the
model based on local CR measurements
and inconsistent with the EGRET GeV excess.
Although the \fermilat{} spectral shape is
consistent with the model, the overall emission in the model predictions using \galprop{} was systematically low
by 10-20\%. This calculation employed an \emph{a priori} model of the diffuse emission, the ``conventional" model~\cite{SMR2000,SMR2004}, that is
based on local CR measurements taken before the \fermilat{} launch.
More detailed studies of molecular clouds in the 2nd and 3rd Galactic quadrants~\cite{Abdo2010cas,Abdo2009loc}
show that the CR proton spectrum does not fluctuate significantly over a large Galactic volume, which supports the
reasoning to use the conventional model based on local CR measurements.
A comparison between the models and the \fermilat{} and INTEGRAL data in the inner Galaxy is discussed in \cite{Strong2011}.

The diffuse Galactic emission presents a strong foreground signal to the much
fainter diffuse extragalactic emission, which is often referred to as the extragalactic
\gray{} background (EGB) and generally assumed to have an isotropic
or nearly isotropic distribution on the sky. 
The EGB is composed of contributions from unresolved
extragalactic sources as well as truly diffuse emission processes, such as
possible signatures of large-scale structure formation,
the annihilation or decay of DM, and many other
processes~\cite{Dermer2007}. 

The \fermilat{} measurement
of the spectrum of isotropic diffuse \gray{} emission from 200 MeV
to 100 GeV is described in~\cite{Abdo2010egb}. 
The isotropic background was found using a simultaneous fit
of the diffuse Galactic \gray{} emission as modeled using \galprop, resolved sources
from the internal \fermilat{} 9-month source list (using the individual localizations
but leaving the fluxes in each energy bin to be separately fitted for each
source), and a model for the solar IC \gray{} emission~\cite{Moskalenko2006,Orlando2007,Orlando2008}. 
The derived EGB spectrum is
a featureless power law with
index $2.41\pm0.05$ and intensity $I(> 100\ {\rm MeV}) = (1.03\pm0.17)\times10^{-5}$ cm$^{-2}$ s$^{-1}$ sr$^{-1}$, 
significantly softer than the one obtained from EGRET
observations~\cite{Sreekumar1998}.
Note that below 2 GeV the \fermilat{} spectrum is in agreement with the spectrum found from the 
\emph{reanalysis} of the EGRET data~\cite{SMR2004egb} which was also based on \galprop{}.
Using the \fermilat-derived EGB, it was possible to set upper limits on
the \gray{} flux from cosmological annihilation of DM~\cite{Abdo2010dm}.

\subsection{Global CR-related luminosity of the Milky Way} 
%#########################################################################
Observations of the diffuse \gray{} emission from normal galaxies (LMC, SMC, M~31) and the starburst galaxies (M~82, NGC~253)
by the \fermilat~\cite{Abdo2010lmc,Abdo2010smc,Abdo2010m82,Abdo2010m31} and by the atmospheric Cherenkov telescopes~\cite{Acciari2009,Acero2009} show that
CRs is a widespread phenomenon associated with the process of star formation. 
The Milky Way is the best-studied non-AGN dominated star-forming 
galaxy, and the only galaxy that direct measurements of CR intensities and 
spectra are available.
However, because of our position inside, the derivation of 
global properties is not straightforward and requires detailed 
models of the spatial distribution of the emission. 
Meanwhile, understanding the global energy budget of processes related to the injection
and propagation of CRs, and how the energy is distributed across the 
electromagnetic spectrum, is essential to interpret the 
radio/far-infrared relation~\cite{Helou1985,Murphy2006}, galactic 
calorimetry~\cite{Volk1989}, and predictions of 
extragalactic backgrounds~\cite{Thompson2007,Murphy2008,Fields2010}, and for
many other studies.

\begin{figure}[tb!]
\centering
\includegraphics[width=.8\linewidth]{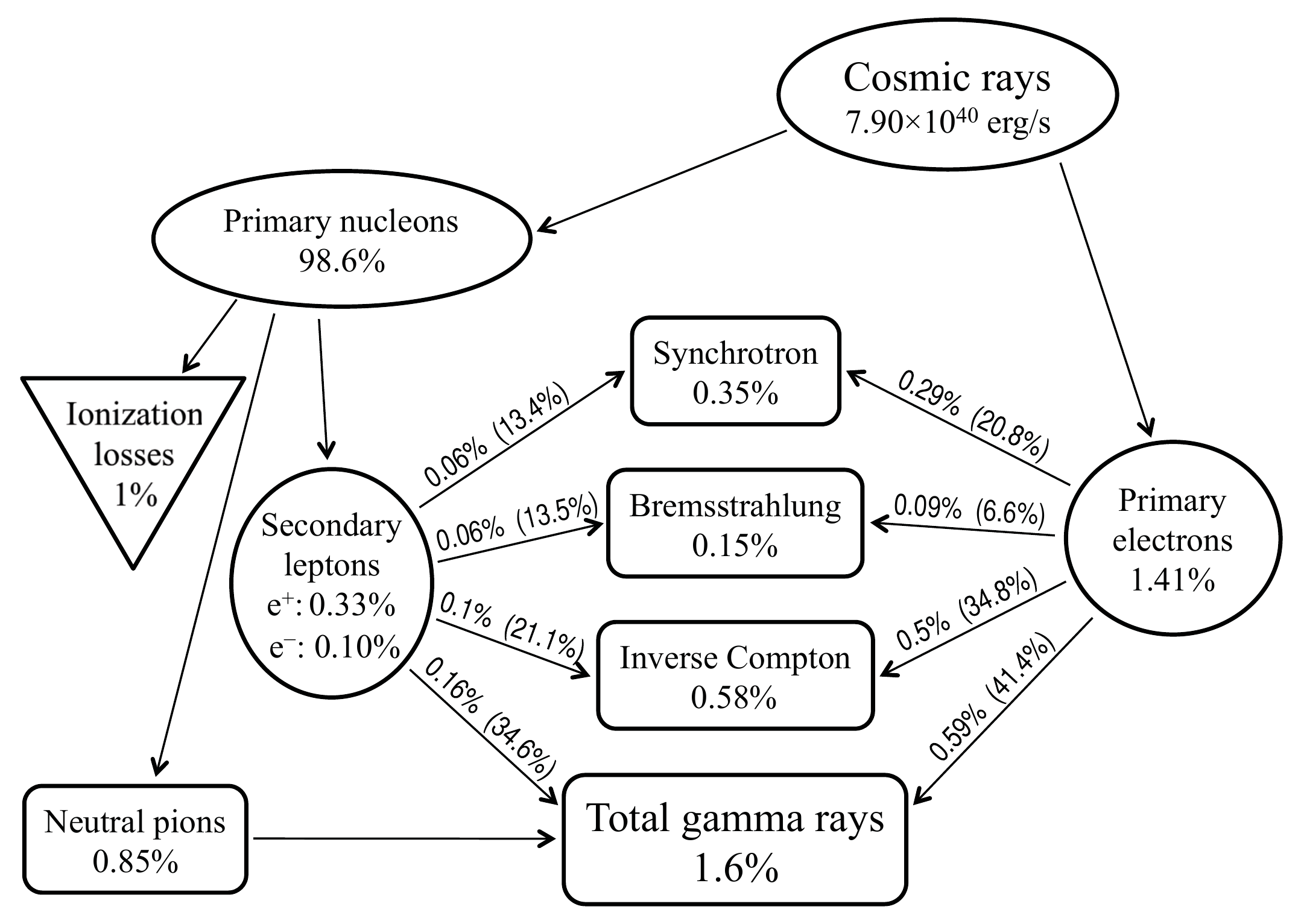}
\caption{The luminosity budget of the Milky Way galaxy 
calculated for a model with 4 kpc halo~\cite{Strong2010}. The percentage figures
are shown with respect to the total injected 
luminosity in CRs. 
The percentages in brackets show the values relative to the luminosity of 
their respective lepton populations 
(primary electrons, secondary electrons/positrons).}
\label{MW}
\end{figure}

Such calculations were carried out in~\cite{Strong2010}. The luminosity spectra were
calculated for representative Galactic propagation models that are 
consistent with
CR, radio, and \gray{} data.
Figure~\ref{MW}
shows the detailed energy budget for a model corresponding to the middle
range of the plausible models.
About 1.8\% of the total CR luminosity goes into the primary and secondary
electrons and positrons, however, the IC scattering contributes half of the total \gray{} luminosity
with the $\pi^0$-decay contributing another half.
The relationship between far-infrared and radio luminosity appears to be consistent with that found for galaxies in general.
The Galaxy is found to be nearly a CR electron calorimeter, 
but {\it only} if \gray{} emitting processes are taken into account. 
The synchrotron emission alone accounts for only one third of the total 
electron energy losses with $\sim$10-20\% of the total synchrotron 
emission from secondary CR electrons and positrons.

\begin{figure}[tb!]
\centering
\includegraphics[width=0.92\linewidth]{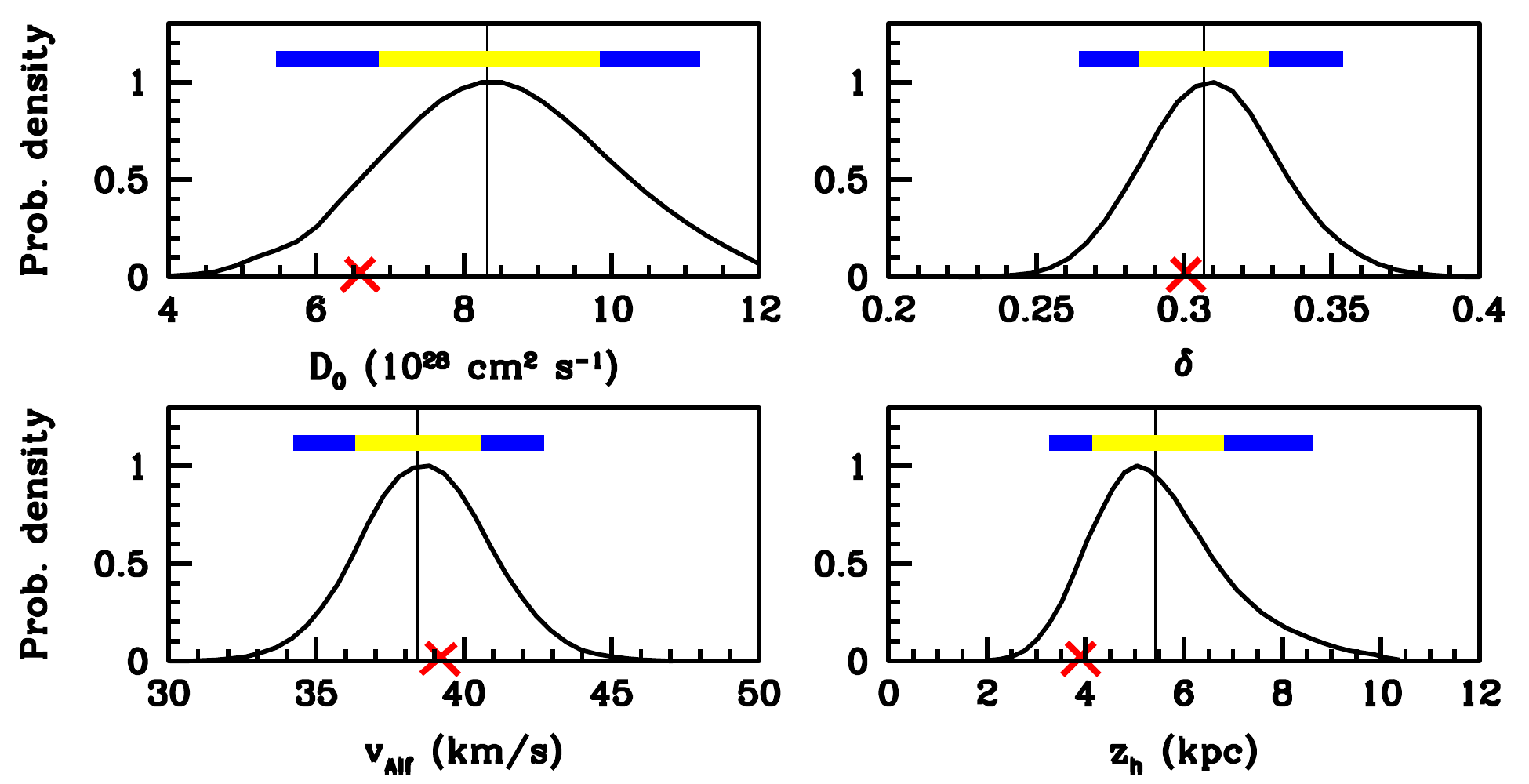}
\caption{1D marginalized posterior probability distribution function (PDF) normalized to the peak for the 
diffusion model parameters, with uniform priors assumed over the parameter ranges~\cite{Trotta2011}. 
The cross represents the best fit, the vertical thin line the 
posterior mean, and the horizontal bar the 68\% and 95\% error
ranges, respectively.}
\label{1D_plots}
\end{figure}

\subsection{Constraints on CR propagation models from a global Bayesian analysis} 
%#########################################################################
The fully Bayesian approach to the problem of deriving constraints for CR propagation models parameters allows one to carry out a global statistical analysis of the whole parameter space, rather than be limited to scanning a reduced number of dimensions at the time. This is important in order 
to be able to fit simultaneously all relevant CR parameters and to explore degeneracies.
While very detailed numerical models of CR propagation exist, a quantitative 
statistical analysis of such models has been so far hampered by the 
large computational effort that those models require. 
Although statistical 
analyses have been carried out before using semi-analytical 
models, the evaluation of the 
results obtained from such models is difficult, as they necessarily 
suffer from many simplifying assumptions.

A full Bayesian parameter estimation has been recently shown to work with a 
\emph{numerical} CR propagation model~\cite{Trotta2011}. 
Despite the heavy 
computational demands of a numerical propagation code, such as \galprop, 
a full Bayesian analysis is possible using 
nested sampling and Markov Chain Monte Carlo methods (implemented in 
the SuperBayeS code~\cite{Ruiz2006,Trotta2008a}).
A remarkable agreement was
found between the ``by-eye'' fitting 
in the past~\cite{SM1998,Moskalenko2002,Ptuskin2006,SM2001a} 
and the parameter constraints from the refined Bayesian inference analysis (Figure~\ref{1D_plots})~\cite{Trotta2011}. 
The posterior mean values of the diffusion 
coefficient $D_0=(8.32 \pm 1.46) \times10^{28}$ cm$^2$ s$^{-1}$ at 4 GV and the
Alfv\'en speed $\valf=38 .4\pm 2.1$ km s$^{-1}$ are in fair agreement 
with earlier estimates  of $5.73\times10^{28}$ cm$^2$ s$^{-1}$ and 
36 km s$^{-1}$~\cite{Ptuskin2006}, respectively. 
The posterior mean halo size is $5.4 \pm 1.4$  kpc, also in agreement with our
earlier estimated range $z_h=4-6$ kpc~\cite{SM2001a}, although our best-fit 
value of $z_h = 3.9$ kpc is somewhat lower, due to the degeneracy 
between $D_0$ and $z_h$. 
However, the well-defined posterior intervals produced in that study are 
significantly more valuable than just the best-fit values themselves as they 
provide an estimate of associated 
theoretical uncertainties and may point to a potential inconsistency 
between different types of data.

%\section{Conclusion}
%#########################################################################

\acknowledgments
\galprop{} development is supported through NASA~Grant~No.~NNX09AC15G.

\bibliography{imos,strong,GalpropBayes,luminosity,FermiDiffusePaper2}
\end{document}